# Tailoring the properties of the 2D ferromagnet CrSBr by lanthanide doping

Sourav Dey[†], Dorye L. Esteras[†] and José J. Baldoví*

Instituto de Ciencia Molecular (ICMol), Universidad de Valencia, c/Catedrático José Beltrán. 2, Paterna 46980, Spain
† Both authors contributed equally
E-mail: j.jaime.baldovi@uv.es

**Abstract**

The growing interest in 2D van der Waals (vdW) magnetic materials stems from their unique properties and potential applications in spintronics, magnonics and quantum information technologies. Among them, CrSBr is a semiconductor that stands out, owing to its high Curie temperature ($T_C \sim 146$ K), air stability and tunable electronic and magnetic properties. Here, we present a systematic investigation of the effects of Dy doping (12.5%, 25%, and 50%) on the structural, electronic and magnetic properties of CrSBr monolayer. Our results reveal that Dy incorporation enhances magnetic anisotropy, modulates $T_C$, and can create quasi-1D ferromagnetic chains that arise from competing ferromagnetic and antiferromagnetic interactions. Additionally, we investigate the properties of DySBr, DySI and DySeI monolayers, which are isostructural to the CrSBr. Our results reveal the feasibility of exfoliating them down to the single-layer and the presence of long-range magnetic order at low temperatures, relying on the combination of both weak exchange interactions and large spin-orbit coupling. This work provides insights into tuning the properties of CrSBr through rare earth doping, unlocking new possibilities for advanced applications at the 2D limit.

## 1. Introduction

The discovery of two-dimensional (2D) van der Waals (vdW) magnetic materials[1-6] represents a milestone in condensed matter physics and materials science, unlocking new possibilities for miniaturised devices in spintronics, magnonics, and quantum information processing.[1, 7-10] Among these materials, the magnetic semiconductor CrSBr stands out due to its unique combination of quasi-one-dimensional (quasi-1D) electronic properties, robust air stability, and relatively high critical temperature ($T_C$) of ~145 K, even at the 2D limit.[11-13] In its few-layer form, CrSBr shows A-type antiferromagnetic order, which leads to significant potential for 2D spintronic applications, exhibiting both gate-tunable magnetism and field-dependent magnetotransport.[4, 6, 14] Moreover, the strong coupling between its electronic and magnetic structures provides advantageous features for the development of 2D magnonic devices, enabling strategies such as strain engineering, doping control and chemical modulation of spin waves.[12, 15, 16]

Despite these fascinating possibilities, the functional applications of 2D magnets still remain limited, primarily due to moderate critical temperatures and weak magnetic anisotropy. In this context, the incorporation of lanthanides into the realm of 2D vdW magnetic materials is a research frontier that can provide a series of benefits. Lanthanides offer distinctive magnetic properties due to the unique characteristics of their 4f electrons, including high magnetic moments and strong spin-orbit coupling. Recent experimental and theoretical studies on lanthanide-based vdW materials have revealed promising chemical and physical properties, leading to the exfoliation of the first examples of rare-earth-based 2D magnets.[17-23] For example, gadolinium-based 2D materials, such as $GdI_2$ and $GdTe_3$, have demonstrated outstanding performance in terms of $T_C$ ~ 241 K[24, 25] and charge mobilities beyond 60,000 $cm^2$ $V^{−1}$ $s^{−1}$,[26] respectively, while the magnetic behaviour of the vdW 2D honeycomb magnet $ErBr_3$ is entirely governed by dipolar interactions.[27] On the contrary, the inner character of 4f orbitals results in weak magnetic exchange interactions, which will lead to a very small $T_C$ (in the order of mK) in most cases. The combination of transition metals and lanthanides may overcome these limitations, leveraging magnetic exchange from transition metals and strong spin-orbit coupling from lanthanides.

A promising approach to achieve this synergy is the addition of lanthanide dopants into transition metal-based 2D magnets. Doping plays a crucial role in functionalising 2D materials by modulating their intrinsic properties.[26] It has been demonstrated that doping transition metals such as Fe, Co, and Ni into 2D transition-metal dichalcogenides induces room-temperature ferromagnetism.[28] Similarly, lanthanide doping has been shown to enhance the catalytic and magnetic characteristics of various 2D systems. For example, Gd-doped $Bi_2MoO_6$ monolayers exhibit increased surface area and catalytic efficiency for nitrogen reduction, resulting in efficient conversion of $N_2$ to $NH_3$,[29] while Gd-doped cerium oxide can be exploited for several applications.[30] On the other hand, Gd-doped $MoS_2$ and $WSe_2$ display enhanced photoresponsivity due to the broader depletion width of the lanthanides,[31] and p-type semiconducting behaviour and ferromagnetism have been observed in $MoS_2$.[32] Furthermore, Gd doping in MXenes such as $Ti_3C_2$ enhances ferromagnetism, which shows a near-room temperature hysteresis loop.[33] All these examples underscore the potential of lanthanide doping in 2D materials for spintronic applications. Notably, some of the strongest permanent magnets combine lanthanides and transition metals, such as NdFeB and SmCo$_5$,[34] evidencing the benefits of this approach. A recent first-principles study demonstrated that rare-earth doping in CrI$_3$ monolayer leads to exciting electronic and magnetic properties and, more importantly, the enhancement of Tc to room temperature.[35]

In this work, we perform a systematic investigation of the effects of Dy at different doping concentrations on the electronic structure and magnetic properties of CrSBr. Our first principles calculations explore doping concentrations of 12.5%, 25% and 50% to evaluate structural modifications, electronic band structure, and magnetic anisotropy of the doped systems. Additionally, we examine the isostructural Dy-based analogues DySBr, DySI and DySeI, whose chemical structures have previously been reported,[36,37] extending the scope of our study to rare-earth-based 2D magnets. This study aims to provide new avenues for optimising CrSBr and other 2D magnets using rare earth elements such as Dy, broadening their application potential in spintronics and magnonics.

## 2. Results and Discussions

Bulk CrSBr crystallises in an orthorhombic FeOCl-like structure (space group Pmmn) (see Figure 1) with lattice parameters $a$ = 3.508 Å, $b$ = 4.763 Å and $c$ = 7.959 Å.[4,14] Within each layer, the Cr atoms reside in distorted octahedra of sulphur and bromine atoms, forming nearly orthogonal Cr–S–Cr and Cr–Br–Cr bonds along $a$ axis and Cr–S–Cr bonds along the $b$ axis with an angle of 162°. This leads to decoupled quasi-1D chains as probed by conductivity measurements.[38] The layers are separated vertically by a vdW gap, allowing the crystal to be exfoliated to the monolayer limit.

In order to explore systematically the effect of Dy-doping on the structural, electronic and magnetic properties of single-layer CrSBr while keeping a reasonable number of atoms in the unit cell, we considered three cases: 12.5% to 25% and 50% of doping. Dy atoms were selected as dopants due to the large magnetic anisotropy of Dy, compared with other lanthanides. We performed first principles based on Hubbard-corrected density functional theory (DFT+U) (see Computational Details) on 2x2 supercells of CrSBr monolayer to consider different dopped structures in a common reference system (Fig. 1(c–f)). The substitution of Cr by Dy elongates the in-plane lattice parameters up to 6.5 and 8.1% for $a$ and $b$, respectively, considering 50% of Dy doping (Table S1). This is due to the larger ionic radius of Dy$^{3+}$ compared to Cr$^{3+}$ and the ionic nature of the 4f orbitals, which increases the Dy-ligand bond length. On the other hand, the out-of-plane $c$-axis remains stable, as it primarily depends on halide composition (see Table S1).

Then, we investigated the magnetic ground state of each doped system using the SpinW code.[39] First, a tight-binding Hamiltonian was derived based on maximally localised Wannier functions (MLWF) using the Wannier90 code.[40] Next, magnetic exchange interactions were computed using the Heisenberg spin Hamiltonian within the TB2J package.[41] Finally, the magnetic configurations were optimised by iteratively rotating the spins in SpinW. The resulting magnetic ground state configurations of CrSBr at 12.5%, 25%, and 50% Dy doping are presented in Figures 1c–f. We also calculated the energies of FM and AFM spin configurations with Dy doping to validate these results and obtained the same magnetic ground state as observed with SpinW (see Fig. S1-3 and Table S2). The results show that AFM coupling between Dy and Cr atoms is favoured, where FM interactions between Cr atoms persist up to 25% Dy doping (Figs. 1d and 1e). At 50% Dy doping, competing magnetic exchange interactions give rise to a more complex magnetic structure, characterised by AFM coupling between Dy and Cr, FM interactions between Cr atoms along the a-axis, and AFM interactions between Cr atoms along the b-axis (Fig. 1f). This suggests the emergence of 1D FM chains coupled antiferromagnetically with one another.

To obtain further insights into the magnetic order, we examined the magnetic moments of the doped systems, which are estimated to be ca. 5.0 $\mu_B$ for Dy and -2.9 $\mu_B$ for Cr, corresponding to the expected spin values of $S = 5/2$ and $S = 3/2$, respectively (Table S3). The S and Br atoms are spin-polarised, with their polarisation decreasing by 67% and 46%, respectively, at 50% Dy doping compared to pristine CrSBr (Table S3). The spin density of these materials is reported in Figure S4, which illustrates that the AFM spin polarisation is highly localised around the Dy atoms. We can also observe that the average magnetic moment of Cr increases and S decreases with doping. This is due to a decrease in covalency, as described in the earlier section (Table S3). Subsequently, we performed a Bader charge analysis to investigate the effect of doping on the charge transfer. Figure S5 and Table S4 indicate that charge flows from metals to ligands due to the latter's higher electron affinity. Compared to CrSBr, the accumulation of charge in the ligands is larger in the doped material (24% and 32% for S and Br, respectively, considering 50% of Dy doping). This is in accordance with the trend observed in the optimised bond distance, which shows the charge transfer is inversely proportional to the length of the metal-ligand bonds.

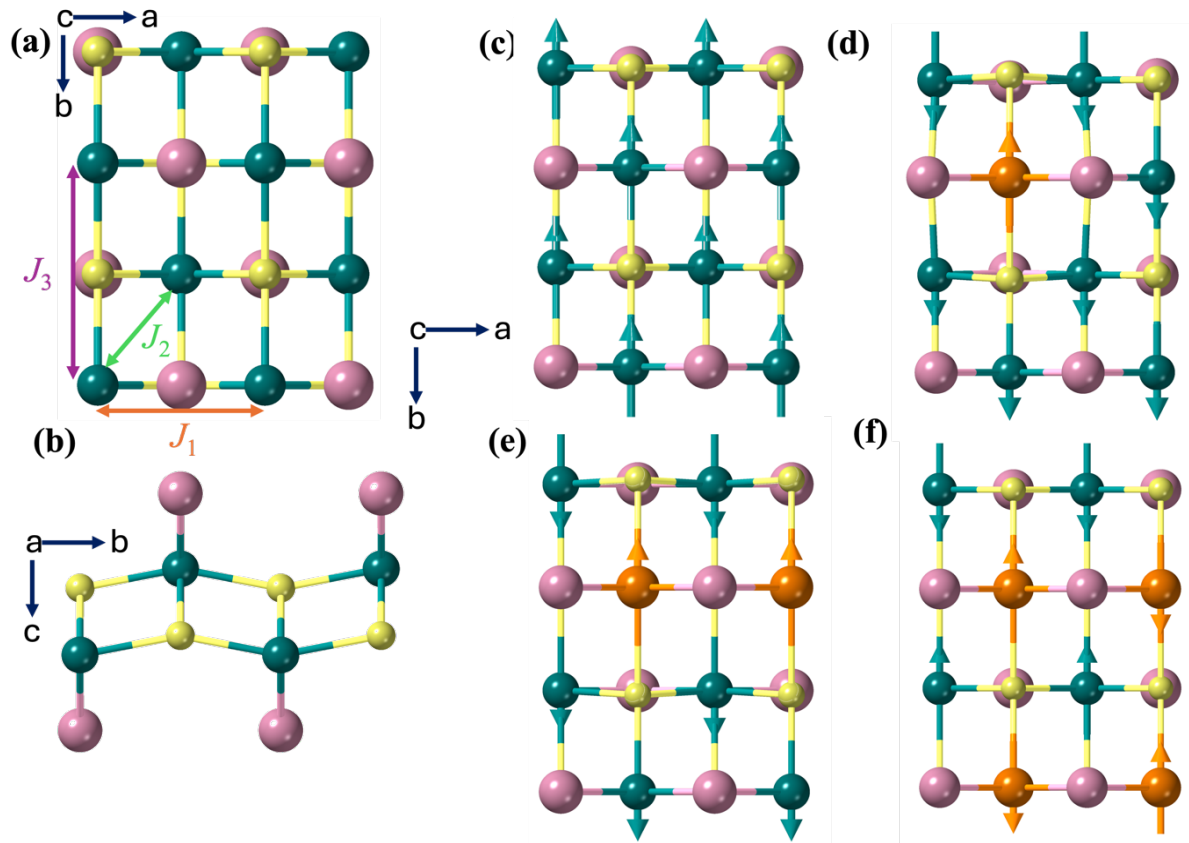

Figure 1: Top and side view of CrSBr monolayer. Colour code: Cr (cyan), Br (pink), S (lime), Dy (orange). The magnetic ground state of (c) CrSBr (d) 12.5% (e) 25% and (f) 50% Dy dop CrSBr. The arrows represent up and down spins, respectively.

For the most stable magnetic configuration, we computed the electronic band structure (Fig. 2) and projected density of states (PDOS) (Figs. S6-S8) for CrSBr at different doping levels. We can observe that at low levels of doping, the valence band maximum (VBM) and conduction band minimum (CBM) are mainly composed of 2p orbitals of S and Cr 3d orbitals, respectively (see Fig. 2 and S6-8). This is similar to pristine CrSBr. The strong mixing between the 3d orbitals of Cr and the ligands, as seen in the PDOS, is consistent with covalent bonding characteristics. In contrast, the absence of such mixing for the 4f orbitals of the lanthanides (like Dy) highlights the more localised nature of these orbitals, reinforcing the idea that the lanthanide-ligand bond is largely ionic. Importantly, Figure 2a shows that the characteristic dispersive conduction bands of CrSBr, which correspond to the d orbitals of Cr, vanish with the partial substitution of Cr by Dy. The progressive filling of the d orbitals displaces these dispersive bands to lower energies, initially closing the band gap (Fig. 2b-c) but producing a large increase of the band gap at large doping (Fig. 2d). As a consequence, the band gap follows an irregular tendency, i.e. 0.59, 0.51, 1.07 eV, with 12.5, 25 and 50 % doping, respectively.

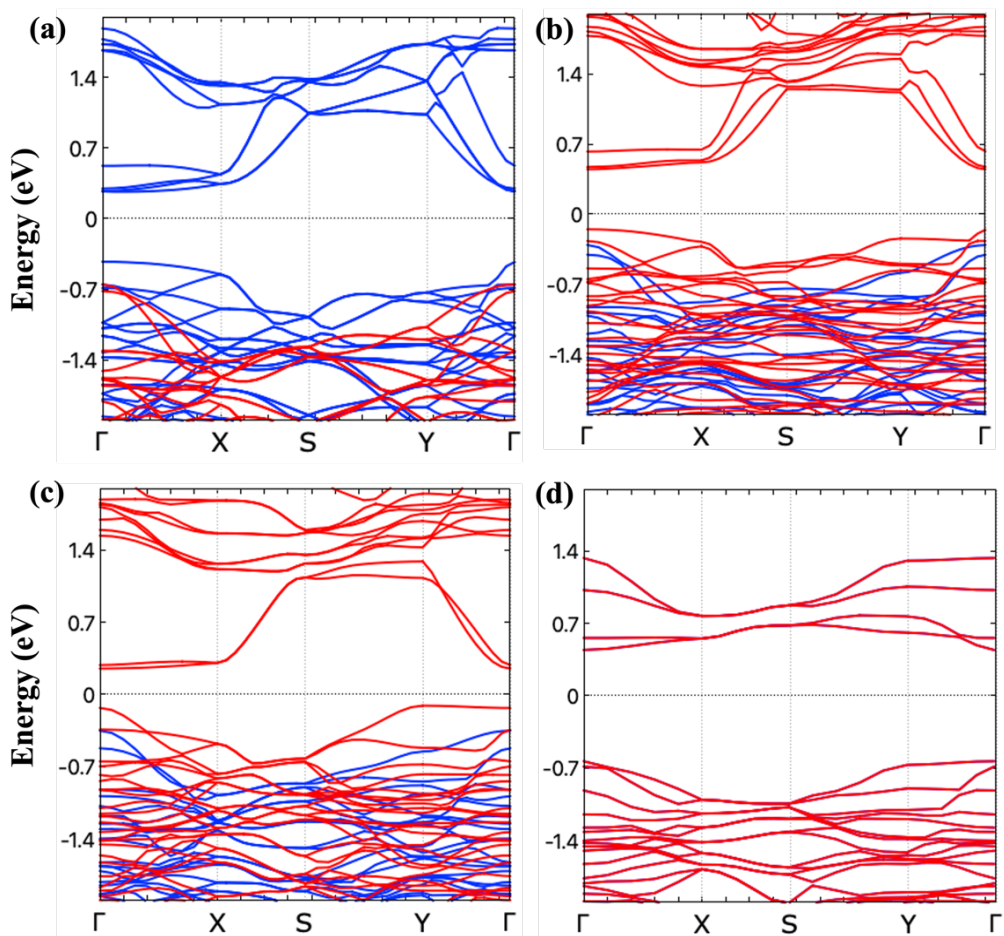

Figure 2: The electronic band structure of 2×2×1 supercell of (a) CrSBr (b) 12.5% (c) 25% and (d) 50% dop CrSBr. Here, the blue and red lines represent the up and down spin, respectively.

The magnetic anisotropy on both metal centres was calculated using multireference CASSCF/RASSI-SO/SINGLE_ANISO calculations using MOLCAS8.2 (see methods and ESI for more details).[42] The computed axial single-ion anisotropy (D for Cr and $3B_2^0$ for Dy) is shown in Table 2 (see also Table S5-8 for more details). The large single-ion anisotropy of Dy compared to Cr reiterates the strong spin-orbit coupling of rare earths compared to transition metals. The change in the sign of D of Cr can be ascribed to the orientation of the magnetic anisotropy axis of Cr, which is pointed along the *b*-axis at 12.5% Dy doping, similar to the undoped CrSBr (Fig. S9), while the axis shifts to an out-of-plane orientation, aligning along the *c*-axis (Fig. S10-S11) with higher levels of doping. The electron density of Dy(III) is oblate in nature; therefore, the increase of the axial ligand field and the decrease of the equatorial ligand field should increase the single-ion anisotropy. Compared to 12.5% Dy doping, the weakening of the metal-ligand bond decreases the equatorial ligand field, therefore increasing the single ion anisotropy of Dy in 25% Dy-dop CrSBr. However, the magnetic anisotropy axis of Dy is oriented along the Dy-Br bond with all levels of doping (Fig. S7-S9).

In the CrSBr monolayer, there are mainly three magnetic exchange interactions between the nearest-neighbours Cr atoms due to superexchange mechanisms through the p orbitals of the ligands. These are labelled as $J_1$, $J_2$ and $J_3$. $J_1$ accounts for the Cr-S-Cr and Cr-Br-Cr interactions along the *a* axis, $J_2$ for the interaction between the nearest Cr atoms placed in two different planes and $J_3$ for the interaction between Cr-S-Cr along the *b* axis (Figure 3a). These exchange interactions are significantly affected by doping as well. In particular, the introduction of Dy atoms substitutes the FM Cr-Cr interactions with AFM and very weak Dy-Cr interactions. As a consequence, the doping of CrSBr inhibits interactions in the material, reducing the number of FM pathways and, at the same time, enhancing FM in the remaining Cr-Cr pairs. This introduces a scenario which requires considering many different neighbour interactions in the spin Hamiltonian (Figs. 3b-d).

The exchange interactions in 12.5% Dy doping can be properly modelled considering eleven nearest neighbour superexchange pathways. From all of them, eight different interactions occur between Cr atoms, referred to as $J_{\text{Cr-Cr}}$, and three occur between Cr and Dy atoms, referred to as $J_{\text{Cr-Dy}}$ (see Fig. 3 and Table 1). Higher levels of doping produce a more uniform situation, being able to simplify the model to eight and, finally, five different types of interactions for 25 and 50 % concentration of Dy, respectively. In particular, for 25 and 50% doping, ultraweak interactions between Dy atoms, $J_{\text{Dy-Dy}}$, are introduced. Generally, the $J_{\text{Cr-Cr}}$ are found to be five times larger than $J_{\text{Cr-Dy}}$ due to the deeply buried nature of the 4f orbitals. The exchange along *a* axis such as $J_{1(\text{Cr-Cr})}$ (4.820 meV) and $J_{3(\text{Cr-Cr})}$ (5.343 meV) are found to be ca. 1 meV larger in 12.5% doping compared to pristine CrSBr. On the other hand, the exchange along the *c* axis, such as $J_{7(\text{Cr-Cr})}$ (3.067 meV) and $J_{8(\text{Cr-Cr})}$ (3.222 meV), are ca. 1 meV smaller than in CrSBr. The exchange interactions between the *a* and *c* axes exhibit two opposing effects compared to CrSBr: $J_{2(\text{Cr-Cr})}$ (4.048 meV) and $J_{5(\text{Cr-Cr})}$ (4.318 meV) increase, while $J_{4(\text{Cr-Cr})}$ (2.727 meV) decreases. To understand the orbitals involved in the magnetic exchange and how it affects the magnitude, we have performed an orbital resolved analysis of magnetic exchange using Green's function approach with TB2J (see computational details).[41] In an octahedral crystal field, the 3d orbitals of Cr splits into two sets, namely $t_{2g}$ ($d_{xy}$, $d_{xz}$ and $d_{yz}$) and $e_g$ ($d_{z^2}$ and $d_{x^2-y^2}$), and 4f orbitals split into three sets, namely, $A_{2u}$ ($f_{xyz}$), $T_{1u}$ ($f_{z^3}$, $f_{y(3x^2-y^2)}$, $f_{z(x^2-3y^2)}$) and $T_{2u}$ ($f_{xz^2}$, $f_{yz^2}$ and $f_{xyz}$). The orbital resolved exchange analysis reveals that $J_{\text{Cr-Cr}}$ interactions are dominated by $t_{2g}$-$e_g$ and $e_g$-$e_g$ pathways, where the former leads to ferromagnetic interactions while the latter leads to antiferromagnetic interactions (see Tables S9-S31 in the ESI). The $J_{\text{Cr-Dy}}$ is found to have two competitive interactions: $t_{2g}$-$T_{1u}$, which contributes to the antiferromagnetic coupling and $e_g$-$T_{1u}$, which contributes to the ferromagnetic coupling (Tables S9-S31). Both of these channels present quite limited overlapping, resulting in a blocking of the interactions between Cr and Dy. Additionally, $J_{\text{Cr-Dy}}$ becomes weaker under larger doping levels due to increased lattice parameters, which leads to greater separation between the metal atoms. This natural tendency induced by the doping results in a continuous inhibition of the critical temperature that is progressively being reduced (128, 78 and 1 K) with the addition of Dy atoms (Fig. S12).

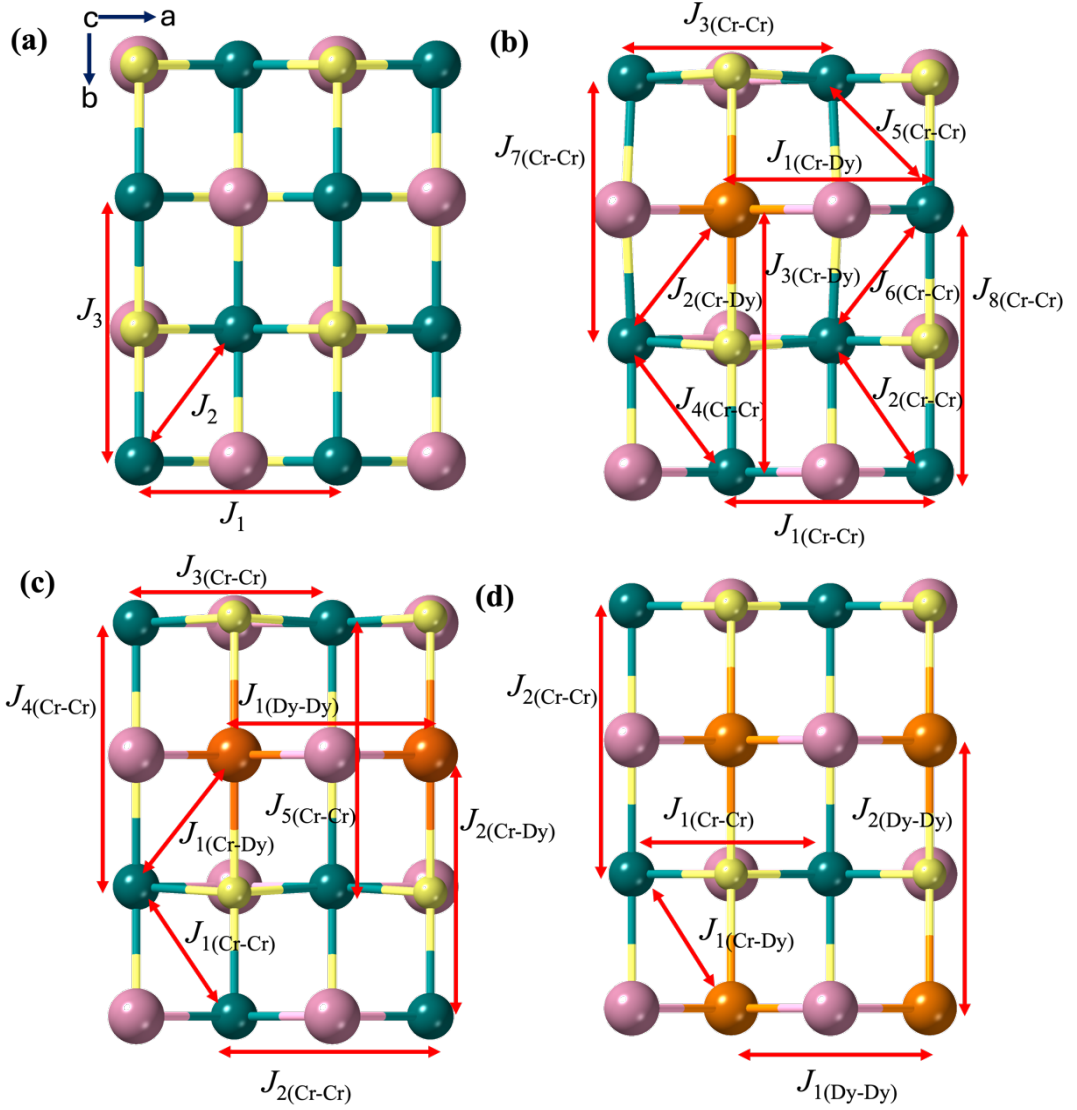

Figure 3: The pathways of magnetic exchange (a) 12.5% (b) 25% and (c) 50% Dy doped CrSBr. See Figure 1 for colour code. (d) The change in curie temperature with the percentage of doping CrSBr.

Table 1: The computed magnetic exchange (in meV, $\hat{H} = -J\hat{S}_1\hat{S}_2$) and single-ion anisotropy for pristine CrSBr and doped CrSBr (see Figure 3 for the labels of exchange).

| CrSBr | | 12.5% of doping | | 25% of doping | | 50% of doping | |
|---|---|---|---|---|---|---|---|
| $J_1$ | 4.17 | $J_{1(Cr-Cr)}$ | 4.820 | $J_{1(Cr-Cr)}$ | 3.864 | $J_{1(Dy-Dy)}$ | -0.172 |
| $J_2$ | 3.41 | $J_{2(Cr-Cr)}$ | 4.048 | $J_{1(Dy-Dy)}$ | -0.115 | $J_{1(Cr-Cr)}$ | 5.347 |
| $J_3$ | 4.57 | $J_{3(Cr-Cr)}$ | 5.343 | $J_{2(Cr-Cr)}$ | 6.401 | $J_{1(Cr-Dy)}$ | -0.251 |
| | | $J_{4(Cr-Cr)}$ | 2.727 | $J_{3(Cr-Cr)}$ | 4.952 | $J_{2(Dy-Dy)}$ | -0.088 |
| | | $J_{1(Cr-Dy)}$ | -0.912 | $J_{1(Cr-Dy)}$ | -0.415 | $J_{2(Cr-Cr)}$ | -1.348 |
| | | $J_{5(Cr-Cr)}$ | 4.318 | $J_{4(Cr-Cr)}$ | 1.256 | | |
| | | $J_{6(Cr-Cr)}$ | 3.380 | $J_{2(Cr-Dy)}$ | -0.217 | | |
| | | $J_{2(Cr-Dy)}$ | -0.588 | $J_{5(Cr-Cr)}$ | 0.280 | | |
| | | $J_{7(Cr-Cr)}$ | 3.067 | | | | |
| | | $J_{8(Cr-Cr)}$ | 3.222 | | | | |
| | | $J_{3(Cr-Dy)}$ | -0.103 | | | | |
| | | D(Cr) | 0.04 | D(Cr) | -0.05 | D(Cr) | -0.10 |
| | | $3B_2^0$ (Dy) | -0.35 | $3B_2^0$ (Dy) | -0.54 | $3B_2^0$ (Dy) | -0.38 |

With the obtained set of parameters from the spin Hamiltonian, we computed the magnon dispersions of these materials using the package spinW.[39] Figure 4 presents the magnon dispersions for the different levels of doping, which present a total of 8 different magnon modes in the spectra coming from the supercells utilised for the doping simulations. As a result of the particular optimisation of the spin configurations, these magnon dispersions present minimum energy in the Γ point. Figure 4a presents a situation with a clear majority of Cr and thus presents a spectrum that lies very close to the spectrum of CrSBr, both in energies and shape (compared in the same cell). The continuous introduction of Dy breaks the FM interactions between chromium atoms ($J_{Cr-Cr}$), producing the previously discussed structure based on chains. As a consequence, the magnon modes are starting to behave independently according to the loss of interconnectivity between exchange interactions. This can be observed in Figure 4 at 25% doping, where the magnon modes start to look very similar despite their being situated at different energies. In addition, this figure also shows that the flat line close to 0 meV, which corresponds to the interactions involving the rare earth, comes closer to 0 energy and results in an important opening of the gap induced by the single ion anisotropy of Dy.

Under 50% doping, the connexions between Cr atoms become limited, and as a result, we still have a spectrum compatible with the CrSBr, but that decreases importantly in energy at the same time that all the independent magnon modes become degenerated. In this spectrum, the ground state energy corresponds already to the almost flat band associated with the Dy spectra.

Finally, in the extreme case of 100% Dy, the magnon spectrum is dominated by the previously discussed flat band, which is zoomed in Figure 4. This scenario presents a complete stabilisation of the Dy ground state discussed in the previous doping, which drastically differs from the CrSBr and presents importantly lower energies in the sub-THz, with a quite important anisotropy gap. More importantly, and as it will be discussed in the next section, the magnon spectrum of pristine DySBr and other similar materials explored in the next section present a Dirac-like dispersion.

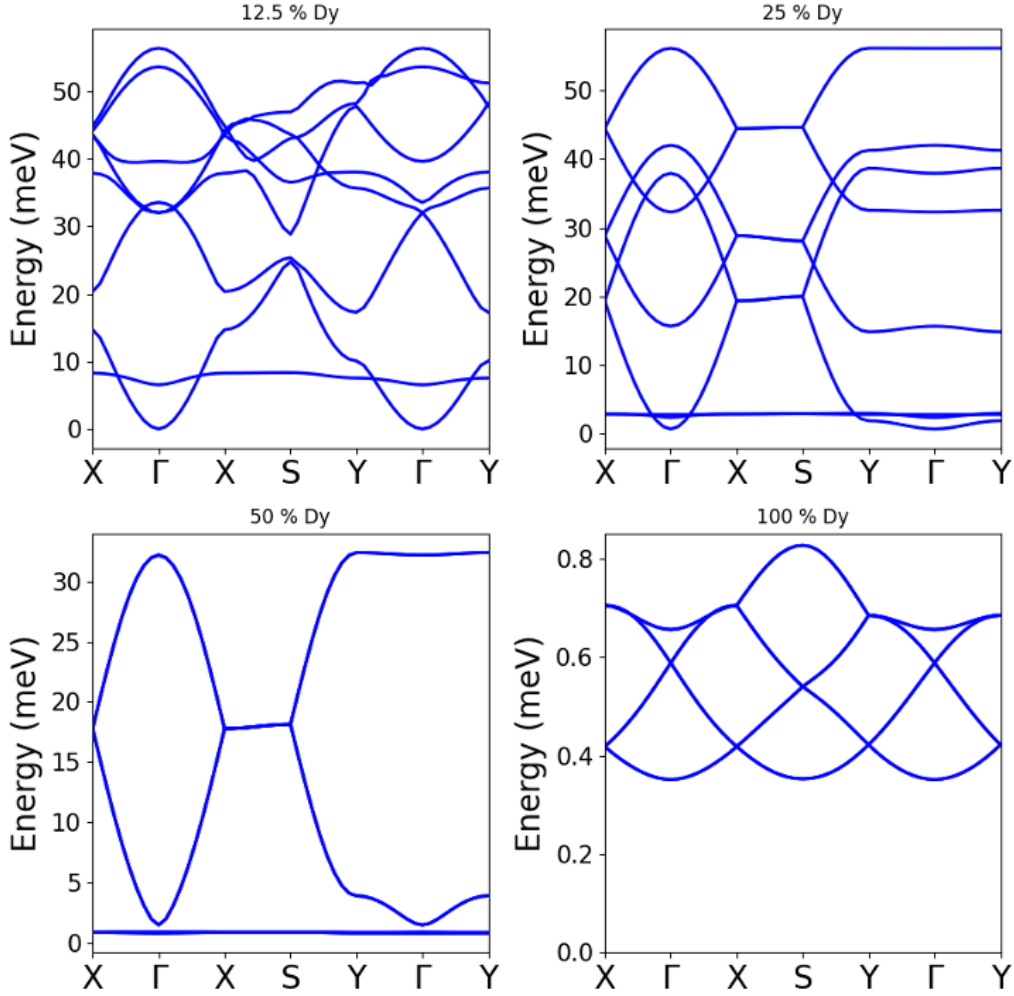

Figure 4: The magnon dispersion of the 2×2×1 supercell of 12.5%, 25%, 50% Dy doped CrSBr and pristine DySBr.

After examining the electronic and magnetic properties of Dy-doped CrSBr, we extended our focus to 100% Dy doping in CrSBr. In this context, we took advantage of lanthanide-based analogues of CrSBr, such as DySBr, DySI, and ErSeI, which were reported two decades ago. [36, 37] These materials share structural similarities with CrSBr, although they have not been magnetically characterised to date and exhibit notable variations in lattice parameters and bond angles. In particular, DySBr features longer in-plane lattice parameters, a (~0.7 Å) and b (~0.6 Å), due to the larger ionic radius of $Dy^{3+}$ compared to $Cr^{3+}$ (Table 2 and Fig. S13). Along the crystallographic axis a, the Dy atoms are connected by S ligands forming Dy–S–Dy bonds with angles of 160° and 158° in DySBr and DySI, respectively. Along *b*, Dy atoms form Dy–S–Dy and Dy–X–Dy bonds with angles of 96(100)° and 93(87)° in DySBr(DySI). As shown in Table 2, a subtle compression of *a* and *c* contrasted with an expansion of *b* is induced by the optimisation of the structure, mirroring the effects seen in CrSBr. In particular, the optimised lattice parameters of DySeI are compared with respect to the originally reported structure of ErSeI, showing small changes as expected given the similar atomic radius between Er and Dy. As a consequence of the elongation of the Dy-chalcogenide bond from sulphur to selenium, DySeI presents overall larger b and c parameters with respect to the other materials, leading to larger Dy–Se–Dy bond angles (167°).

Table 2: The optimised (experimental) lattice parameters in Å of CrSBr, DySBr, DySI and DySeI. The experimental lattice parameters of DySeI correspond to the original structure of ErSeI. (* values correspond to the experimental one for CrSBr)

| Material | a | | b | | c | |
|---|---|---|---|---|---|---|
| | *Bulk* | *Monolayer* | *Bulk* | *Monolayer* | *Bulk* | *Monolayer* |
| CrSBr | 3.508* | 3.549 | 4.763* | 4.754 | 7.959* | 25.965 |
| DySBr | 4.043 (4.075) | 4.035 | 5.412 (5.351) | 9.656 | 7.893 (8.070) | 25.965 |
| DySI | 4.137 (4.179) | 4.135 | 5.395 (5.325) | 9.848 | 8.450 (9.226) | 25.965 |
| DySeI | 4.127 (4.183) | 4.145 | 5.644 (5.584) | 10.278 | 8.790 (8.890) | 25.965 |

To explore the feasibility of exfoliating these materials down to the monolayer limit, we calculated the mechanical cleavage energy by mimicking the exfoliation process (Fig. S14) and increasing the distance between adjacent layers separated by vdW interactions (see computational details). The computed exfoliation energy is estimated to be around 0.18-0.21 J/m$^2$ in DySBr, DySI and DySeI (Fig. S14), which is in the order of other 2D materials such as graphene (0.33 J/m$^2$) or transition metal-based 2D materials such as MnPSe$_3$ (0.23 J/m$^2$), CrI$_3$ (0.30 J/m$^2$) and CrSBr (0.21 J/m$^2$).[43-45] This suggests the possible exfoliation of the family of lanthanide chalcogen halides in the limits of miniaturisation. Thus, we explored the electronic and magnetic properties of these materials in the monolayer phase, which possess very similar lattice parameters to the bulk (Table 2), indicating the 2D nature of these materials.

Our DFT+U simulations predict an antiferromagnetic ground state in all three lanthanide-based 2D materials, which contrasts with CrSBr (see Fig. S15 and Table S32). Nevertheless, the FM order and different AFM arrangements lie close in energy to the ground state (around ca. 0.4-1.2 meV/f.u., see Table S32) because of the weak order fomented by the low-interactive behaviour of the lanthanide atoms. All these compounds are characterised by the presence of the large magnetic moments provided by the five unpaired electrons of the Dy atoms (ca. 5.0 μ$_B$, Table S33 and Fig. S16). The Bader charge analysis suggests that, among the three compounds, the largest charge transfer is observed in DySBr, followed by DySI and DySeI (Table S34 and Fig. S17), suggesting that among these three compounds, DySeI is the most covalent, followed by DySI and DySBr.

The electronic band structure revealed the AFM ground state in these materials as up and down spin states coincide (Fig. S18). In Figure 5, we analyse the AFM ground state and the effect of spin-orbit coupling (SOC) on the features around their bands. The effect of SOC in the electronic states of these crystals is noticeable due to the presence of the rare earth element, which leads to energy shifts in the band structure and splitting of the 4f orbitals induced by SOC. The electronic structure of these lanthanide compounds presents direct band gaps of 2.99, 2.56 and 1.90 eV for DySBr, DySI and DySeI, respectively, that, in the absence of SOC, evolve to 2.84, 2.57 and 2.1 eV (Fig. 5 and S19). The rearrangement of the bands induced by the SOC splitting leads to a reduction in the band gap in the case of DySBr. In contrast, DySeI suffers an important gap opening of 0.2 eV, and the DySI band gap remains unaffected as a consequence of the symmetric shift in energies of both valence and conduction bands. The band gap follows a decreasing tendency from DySBr to DySI to DySeI, which is dominated by crystal field splitting

(DySBr > DySI > DySeI).[3, 46-51] Compared to CrSBr, the band gaps of the Dy compounds are ca. 1 eV larger in energy. This is due to the difference in conduction bands: in CrSBr, they are formed by the empty 3d orbitals, which are lower in energy compared to the 5d orbitals of lanthanides that form their conduction bands.[52-54] The orbital character of the bands is described in the projected density of states (PDOS) in Figures S19-S21, which indicates the important presence around the Fermi level of the np orbitals that belong to the halides and chalcogenides. In contrast, the 4f orbitals of Dy are positioned in the occupied low energy bands (5-6 eV below the Fermi) and in the empty states that form the conduction bands. An orbitally resolved analysis of the bands in the presence of SOC (Fig. S19) provides a more detailed analysis of the role of the orbital states in the electronic structure. A close look into Figure S20 shows in the presence of SOC, the band gap in these materials is dictated by the $np_{3/2}$ states from the chalcogenide atoms that form the top of the valence band and the vacant 4f and 5d Dy states that originate at the bottom of the conduction band. The important contribution of the ligands around the Fermi level is enhanced in DySI by the presence of I (dark blue). However, in DySeI, the valence bands close to the Fermi level are dominated by the presence of the Se p orbitals, which displace the effect of the halides. In parallel, the gap originated at 4 eV below the Fermi and is progressively broadening from DySBr to DySeI.

To further explore the effect of SOC on the properties of these materials, we performed calculations of magnetic anisotropy energy (MAE = $E_\parallel - E_\perp$) to unveil the spin orientation and the presence of easy and hard magnetic axes in these compounds. MAE calculations (see methods) carried out with the spins in the different space axes reveal the spins lie in out-of-plane in the case of all the materials. Moreover, as a direct consequence of the large SOC of Dy, MAE results in importantly higher values compared to usual 2D magnets (7.4, 9.9 and 15.4 meV per magnetic atom for DySBr, DySI and DySeI, respectively).[13, 55] On the other hand, these values are in line with rare earth elements.[56, 57] In addition, we consider the contribution of the magnetic dipoles to the magnetic anisotropy, which is quite important in CrSBr, computing shape anisotropy. Despite the significant differences in the axes of these structures, shape anisotropy results to be negligible (~ 0.08 meV) in these systems compared with the important contribution of the SOC-driven magnetic anisotropy, presenting a situation that contrasts with the particular case of CrSBr.

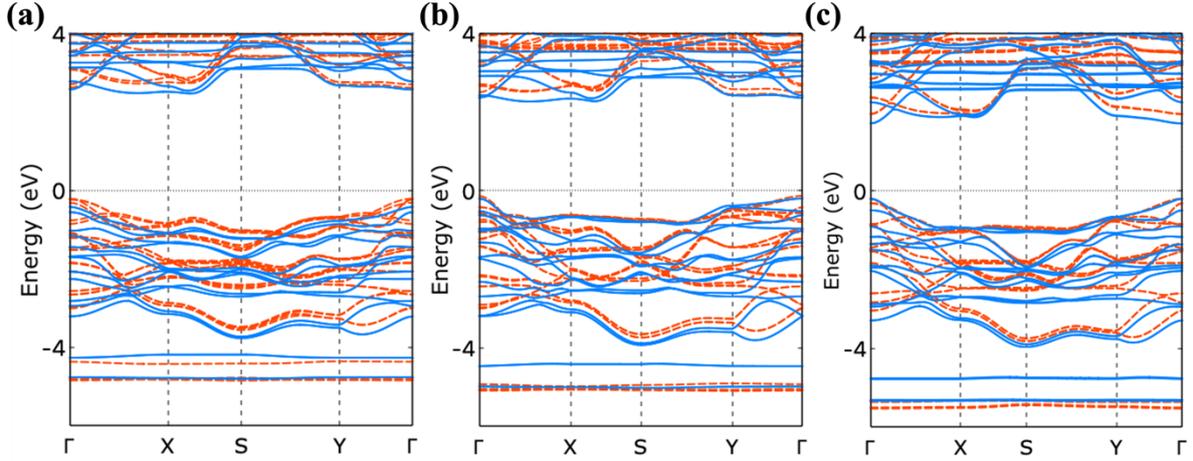

Figure 5: Comparison of DFT+U calculated band structures (monolayer) with and without spin-orbit coupling (SOC) of (a) DySBr, (b) DySI and (c) DySeI

The single-ion anisotropy was found to decrease from DySBr to DySI, and then it increases to DySeI ($3B_2^0$ = -0.20, -0.17, -0.19 meV in DySBr, DySI and DySeI, respectively, see Tables S35-38).[58] The decrease of the axial ligand field from bromide to iodide decreases the single-ion anisotropy from DySBr to DySI. On the other hand, the decrease of the equatorial ligand field from sulphur to selenium increases the single-ion anisotropy from DySI to DySeI. The single ion anisotropy axis was found to be oriented along the out-of-plane Dy-halide bond, which confirms that the easy axis of the system is contained in the c direction. contained along the out-of-plane Dy-halide bond (Fig. S23).[42]

Similar to CrSBr, the exchange interactions in these systems can be modelled by three nearest neighbour isotropic exchange parameters: $J_1$, $J_2$ and $J_3$. The computed magnetic exchange interactions are found to be AFM and very weak (Table 39). The obtained isotropic exchange parameters are found to be one order of magnitude lower than the interactions found in CrSBr due to the non-interactive character of the internal 4f orbitals. The weak exchange interactions formulate an important competition between exchange and dipolar interactions that are of the same order of magnitude. The orbital resolved magnetic exchange indicates that $J_1$ mostly arise from the $T_{1u}$-$T_{1u}$ mechanisms, and their small contribution is similar in all the materials (Tables S40-S48, Figure S24). On the other hand, $J_2$ also takes place through the $T_{1u}$-$T_{1u}$ channels, which follow a decreasing trend from DySBr to DySI to DySeI (Figure S24 and Table S40-S48). Finally, $J_3$ can be described with the $T_{2u}$-$T_{2u}$ interactions that follow the same trend (Figure S24 and Tables S40-S48). As a consequence of the very low exchange interactions, the critical temperature of these materials results to be extremely low (~ 1 K).

Finally, we calculate the magnon dispersion of DySBr, DySI and DySeI (Figure 6). We may observe that the frequency of the magnon modes is of the order of 1 meV; thus, frequencies are contained in the sub-THz regime (hundreds of THz), with a general absence of important differences between them. This range of energies has already been measured experimentally in ErBr$_3$ and points to the viability of measuring magnon transport in these systems.[27] More interestingly, the magnon spectra of these materials present a Dirac-like dispersion in the X point of the Brillouin zone, which motivates the exploration of other lanthanide materials, such as Er analogues, because the in-plane anisotropy produced by the dipoles is expected to be more intense.

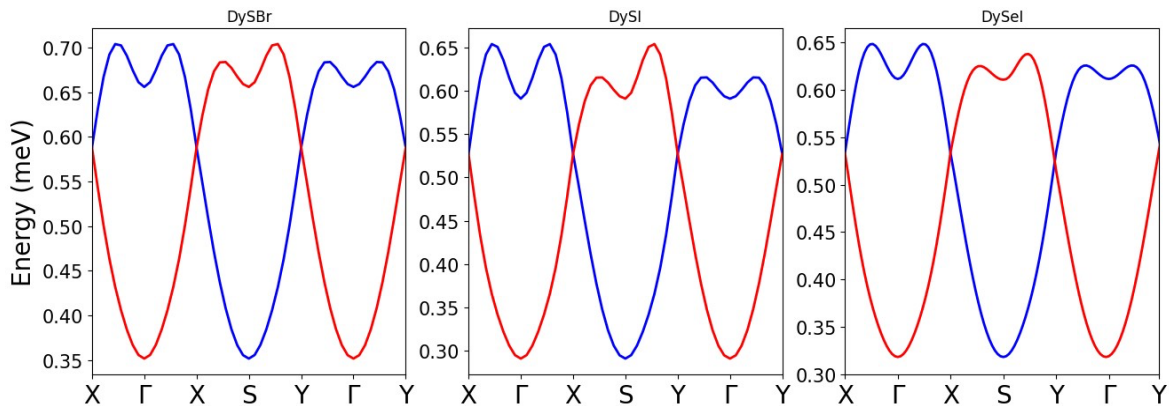

Figure 6. Magnon dispersions in the unit cell for DySBr, DySI and DySeI. Red and blue lines indicate the two different magnon modes.

## 3. Conclusions

In this work, through first-principles calculations, we have investigated the electronic and magnetic properties of Dy-doped CrSBr and its lanthanide-based analogues, DySBr, DySI, and DySeI. Our findings demonstrate that Dy doping significantly alters the magnetic order in CrSBr, enhancing spin anisotropy while introducing competing ferromagnetic and antiferromagnetic interactions. This leads to a complex spin arrangement, including the emergence of quasi-1D ferromagnetic chains that drastically affect the Curie temperature and magnetic properties of CrSBr. Additionally, our analysis of magnon dispersions reveals that Dy doping in CrSBr disrupts the ferromagnetic exchange pathways, having an important impact on the magnon modes that become progressively disconnected from each other as long as the Dy concentration restricts the Cr-Cr interactions. In addition, we highlight the viability of exfoliating lanthanide-based materials to the monolayer limit, which presents cleavage energies comparable to established 2D materials like graphene and $CrI_3$. These new lanthanide 2D materials provide quite interesting magnon spectra in the sub-THz regime, with distinctive Dirac-like dispersions at the X point of the Brillouin zone. These unique magnonic features support the potential interest of future investigation of rare-earth doping and lanthanide-based 2D materials. Moreover, the demonstrated interplay between spin anisotropy, magnetic exchange, and magnonic behaviour underscores the versatility and promise of these systems for 2D-based quantum technologies applications.

## 4. Computational Details

All the electronic structure calculations have been performed using spin-polarised density functional theory (DFT) with the Quantum ESPRESSO package.[59] We have followed a DFT+U approach (U is the on-site coulomb repulsion) using an ortho-atomic U = 8 eV to describe the strong correlation of electrons in rare earth elements.[60-64] The Perdew−Burke−Ernzerhof (PBE) functional and the generalised gradient approximation (GGA) were considered to describe the exchange-correlation energy.[65] Fully relativistic (FR) ultrasoft (US) pseudopotentials (PP) were chosen from the Pslibrary using the code of Andrea Dal Corso to include the spin-orbit coupling of lanthanides.[66] We have employed 155 Ry (100 Ry in pristine CrSBr) and 1550 Ry (800 Ry in pristine CrSBr) cutoffs to describe the wave function and the charge density, respectively (Magnetic anisotropy energy calculations were performed with extended cutoffs of 305 and 3050 Ry for the wave functions and the charge density, respectively). The geometry optimisation of all the structures was performed using the Broyden−Fletcher−Goldfarb−Shanno (BFGS) algorithm until the force on each atom was < $1\times10^{-3}$ Ry/au and the energy difference between two consecutive relaxation steps was <$1\times10^{-4}$ Ry.[67] The Brillouin zone was sampled by a Γ−centred 8×6×1 Monkhorst−Pack mesh for plane-wave calculations.[68] The cleavage energy was estimated by considering a bilayer surrounded by an 18 Å vacuum and systematically varying the interlayer distance (d) with respect to the optimised interlayer distance in the bulk ($d_0$) until convergence. Monolayer calculations were performed, isolating the systems with 18 Å vacuum in the c direction to avoid interlayer interactions with periodic images. The 4f orbitals of the rare earth, 3d orbitals of transition metal and the valence p orbitals of the ligands were chosen as projectors to construct the maximally-connected subspace for Wannier90 calculations, ensuring a good fit to the electronic band structure and highly localised spreads.[40] Exchange interactions were estimated using Green's function method with the TB2J software.[41] We have used the Gd analogue to compute isotropic exchange and also to compute the energy differences between FM and other AFM, as illustrated in Table S2 and Figure S2. This methodology is employed quite often in molecular magnetism due to the weak nature of magnetic exchange in lanthanide complexes.[69,70] Spin waves were simulated using the linear spin wave theory implemented in the software SpinW[39], combining positive and negative solutions according to the Colpa method. The Tc value was computed with Vampire code by performing atomistic simulations with 50000 equilibrium steps and 50000 averaging steps utilizing llg-heun integrator.[71] In this regard, we have selected supercells with dimensions of 22×23×4 for the dopped materials. It is important to note that we have employed calculated single-ion anisotropies of Cr and Dy to compute both the spin wave and Tc values.

**Acknowledgements**
The authors acknowledge the financial support from the European Union (ERC-2021-StG-101042680 2D-SMARTiES, FET-OPEN SINFONIA 964396 and Marie Curie Fellowship SpinPhononHyb2D 10110771), the Spanish MICINN (Excellence Unit "María de Maeztu" CEX2019-000919-M). The computations were performed on the TirantIII cluster of the Servei d'Informàtica of the University of Valencia.